\documentclass[11pt]{article}
\pdfoutput=1
\usepackage{amsfonts,amsmath}
\usepackage{amssymb}
\usepackage{fancyhdr}
\usepackage{slashed}
\usepackage{graphicx}
\usepackage{subfigure}
\usepackage{color}

\usepackage{hyperref}
\usepackage[utf8]{inputenc}
\usepackage[titletoc]{appendix}

\def\hybrid{\topmargin -20pt    \oddsidemargin 0pt
        \headheight 0pt \headsep 0pt
        \textwidth 6.25in       
        \textheight 9.25in       
        \marginparwidth .875in
        \parskip 5pt plus 1pt   \jot = 1.5ex}

\hybrid

\def\baselinestretch{1.2}

\catcode`\@=11

\def\marginnote#1{}
\newcount\hour
\newcount\minute
\newtoks\amorpm
\hour=\time\divide\hour by60
\minute=\time{\multiply\hour by60 \global\advance\minute by-\hour}
\edef\standardtime{{\ifnum\hour<12 \global\amorpm={am}%
        \else\global\amorpm={pm}\advance\hour by-12 \fi
        \ifnum\hour=0 \hour=12 \fi
        \number\hour:\ifnum\minute<10 0\fi\number\minute\the\amorpm}}
\edef\militarytime{\number\hour:\ifnum\minute<10 0\fi\number\minute}

\def\draftlabel#1{{\@bsphack\if@filesw {\let\thepage\relax
   \xdef\@gtempa{\write\@auxout{\string
      \newlabel{#1}{{\@currentlabel}{\thepage}}}}}\@gtempa
   \if@nobreak \ifvmode\nobreak\fi\fi\fi\@esphack}
        \gdef\@eqnlabel{#1}}
\def\@eqnlabel{}
\def\@vacuum{}
\def\draftmarginnote#1{\marginpar{\raggedright\scriptsize\tt#1}}

\def\draft{\oddsidemargin -.5truein
        \def\@oddfoot{\sl preliminary draft \hfil
        \rm\thepage\hfil\sl\today\quad\militarytime}
        \let\@evenfoot\@oddfoot \overfullrule 3pt
        \let\label=\draftlabel
        \let\marginnote=\draftmarginnote
   \def\@eqnnum{(\theequation)\rlap{\kern\marginparsep\tt\@eqnlabel}%
\global\let\@eqnlabel\@vacuum}  }

\def\preprint{\twocolumn\sloppy\flushbottom\parindent 2em
        \leftmargini 2em\leftmarginv .5em\leftmarginvi .5em
        \oddsidemargin -.5in    \evensidemargin -.5in
        \columnsep .4in \footheight 0pt
        \textwidth 10.in        \topmargin  -.4in
        \headheight 12pt \topskip .4in
        \textheight 6.9in \footskip 0pt
        \def\@oddhead{\thepage\hfil\addtocounter{page}{1}\thepage}
        \let\@evenhead\@oddhead \def\@oddfoot{} \def\@evenfoot{} }

\def\numberbysection{\@addtoreset{equation}{section}
        \def\theequation{\thesection.\arabic{equation}}}

\def\underline#1{\relax\ifmmode\@@underline#1\else
        $\@@underline{\hbox{#1}}$\relax\fi}

\def\titlepage{\@restonecolfalse\if@twocolumn\@restonecoltrue\onecolumn
     \else \newpage \fi \thispagestyle{empty}\c@page\z@
        \def\thefootnote{\fnsymbol{footnote}} }

\def\endtitlepage{\if@restonecol\twocolumn \else \newpage \fi
        \def\thefootnote{\arabic{footnote}}
        \setcounter{footnote}{0}}  

\catcode`@=12
\relax

\def\figcap{\section*{Figure Captions\markboth
        {FIGURECAPTIONS}{FIGURECAPTIONS}}\list
        {Figure \arabic{enumi}:\hfill}{\settowidth\labelwidth{Figure
999:}
        \leftmargin\labelwidth
        \advance\leftmargin\labelsep\usecounter{enumi}}}
 \relax
\def\tablecap{\section*{Table Captions\markboth
        {TABLECAPTIONS}{TABLECAPTIONS}}\list
        {Table \arabic{enumi}:\hfill}{\settowidth\labelwidth{Table
999:}
        \leftmargin\labelwidth
        \advance\leftmargin\labelsep\usecounter{enumi}}}
 \relax
\def\reflist{\section*{References\markboth
        {REFLIST}{REFLIST}}\list
        {[\arabic{enumi}]\hfill}{\settowidth\labelwidth{[999]}
        \leftmargin\labelwidth
        \advance\leftmargin\labelsep\usecounter{enumi}}}
 \relax
\makeatletter
\newcounter{pubctr}
\def\publist{\@ifnextchar[{\@publist}{\@@publist}}
\def\@publist[#1]{\list
        {[\arabic{pubctr}]\hfill}{\settowidth\labelwidth{[999]}
        \leftmargin\labelwidth
        \advance\leftmargin\labelsep
        \@nmbrlisttrue\def\@listctr{pubctr}
        \setcounter{pubctr}{#1}\addtocounter{pubctr}{-1}}}
\def\@@publist{\list
        {[\arabic{pubctr}]\hfill}{\settowidth\labelwidth{[999]}
        \leftmargin\labelwidth
        \advance\leftmargin\labelsep
        \@nmbrlisttrue\def\@listctr{pubctr}}}
 \relax
\makeatother
\newskip\humongous \humongous=0pt plus 1000pt minus 1000pt

\newif\ifdtup

\relax

\def\be{\begin{equation}}
\def\ee{\end{equation}}
\def\ba{\begin{eqnarray}}
\def\ea{\end{eqnarray}}

\def\a{\alpha}

\def\g{\gamma}

\def\P{\Pi}

\def\th{\theta}

\def\L{\Lambda}
\def\s{\sigma}

\def\cL{{\cal L}}

\def\cD{{\cal{D} } }

\def\cD{{\cal D}}  
  
  \def\cL{{\cal L}}

\newcommand{\prt}[1]{{\left( {#1} \right)}}
\newcommand{\prtt}[1]{{\left[ {#1} \right]}}

\def\no{\noindent}

\def\IR{\relax{\rm I\kern-.18em R}}

\def\pp{\partial}

\newcommand{\ff}{\frac}

\def\IR{\relax{\rm I\kern-.18em R}}
\def\IL{\relax{\rm I\kern-.18em L}}

\def\inv{^{\raise.15ex\hbox{${\scriptscriptstyle -}$}\kern-.05em 1}}

\def\cL{{\cal L}}

\def\bea{\begin{eqnarray}}
\def\eea{\end{eqnarray}}
\newcommand{\eq}[1]{(\ref{#1})}
\def\nn{\nonumber}

\newcommand{\la}[1]{\label{#1}}

\def\a{\alpha}      
       
\def\g{\gamma}

 \def\L{\Lambda}

 \def\P{\Pi}

\def\s{\sigma}  

\def\th{\theta}

\definecolor{markcolor2}{rgb}{1,0,0}

\definecolor{markcolor3}{rgb}{0,1,0}

\begin{document}

\renewcommand{\theequation}{\thesection.\arabic{equation}}
\csname @addtoreset\endcsname{equation}{section}

\newcommand{\beq}{\begin{equation}}
\newcommand{\eeq}[1]{\label{#1}\end{equation}}
\newcommand{\ber}{\begin{eqnarray}}
\newcommand{\eer}[1]{\label{#1}\end{eqnarray}}
\newcommand{\eqn}[1]{(\ref{#1})}
\begin{titlepage}

\begin{center}

~
\vskip 1 cm

{\Large
\bf Velocity Laws for Bound States in Asymptotically AdS  Geometries
}

\vskip 0.7in

 {\bf Dimitrios Giataganas}
 \vskip 0.1in
 {\em
 	${}^1$  Department of Physics,\\ National Sun Yat-sen University,  \\
 	Kaohsiung 80424, Taiwan
 	\vskip .1in
 	${}^2$  Center for Theoretical and Computational Physics, \\
 	National Sun Yat-sen University					
 	\\
 	Kaohsiung 80424, Taiwan\vskip .1in
 	${}^3$ Physics Division, National Center for Theoretical
 		Sciences, \\
 	Taipei 10617, Taiwan
 \\\vskip .15in
 {\tt ~  dimitrios.giataganas@mail.nsysu.edu.tw
 }\\
 }

\vskip .2in
\end{center}

\vskip .8in

\centerline{\bf Abstract}

We study the behavior of heavy quark bound states in moving plasmas that are dual to theories with generic non-trivial renormalization group flows interpolating between an AdS geometry in the ultraviolet and infrared fixed points with broken symmetries. We investigate analytically the observables associated with the bound state and find their scaling exponents with respect to the Lorentz factor for ultrarelativistic motion. Despite having asymptotically  an AdS geometry, the scaling is not universal and depends on geometric conditions of the Fefferman-Graham expansion in the near boundary regime, or equivalently on the order of the asymptotic background expansion that provides the leading contributions to the Wilson loops. 
\no
\end{titlepage}
\vfill
\eject


\noindent


\def\baselinestretch{1.2}
\baselineskip 19 pt
\noindent


\setcounter{equation}{0}

\section{Introduction}

The Wilson loop operators contain important gauge invariant information about non-perturbative physics of gauge field theories. Their expectation value is an order parameter for the confining-deconfining phase transitions, where confinement indicates an area law. Additionally the Wilson loops above the crossover point of the hadronic state, in the finite temperature strongly coupled deconfined plasma state, are related to a number of different observables which have theoretical interest and some of them are in principle accessible in heavy ion experiments.

The expectation values of Wilson loops contain ultraviolet divergences which can be realised in the holographic gravity dual theories from the infinite bulk to boundary proper distance where the Wilson loop surface extends, while in the field theory can be realized for example, as coming from propagators connecting coincident points at loop computations. In holography there is a rigorous renormalization formalism  \cite{Maldacena:1998im,Drukker:1999zq,Chu:2008xg}, established for static Wilson loops, which can be realised with the Legendre transform method, or equivalently with the holographic renormalization, or with the infinite mass subtraction. The latter scheme not only cancels the ultraviolet divergences but contributes certain infrared terms, which can be thought as part of the thermal masses of the infinite heavily bound state when the loops are orthogonal. Therefore the application of the mass renormalization gives the interaction energy of an extremely massive bound state in a strongly coupled environment. Non-static Wilson loops, or boosted static,  have also been studied and can be thought as corresponding to moving bound states in a thermal strongly couple environment. Some of the renormalization schemes have been extended for this case successfully, for example \cite{Liu:2006he,Chu:2016pea}, despite the complications in the application of the mass renormalization due to the interactions of the color charged singlets in strongly coupled environments  \cite{Giataganas:2022qgq}.

It is known that the application of the mass renormalization scheme, leads to a critical value of the renormalized  Wilson loop expectation value, beyond which the state does not exist. For instance, by fixing the scales of the theory, there exists a critical value for the size of the heavy quark  bound state $L$, beyond which the bound state seizes to exist and  becomes energetically non-favourable compared to the free state of its ingredients, the two infinite massive heavy quarks. $L$ naturally depends  on the scales of the theory and the velocity of motion. It has been found holographically that for ultrarelativistic velocities there is a characteristic power that the screening length depends on the Lorentz factor $\g=1/\sqrt{1-v^2}$, which for AdS spacetimes is $1/4$ when expressed in terms of the velocity $L_{max}\sim \prt{1-v^2}^{1/4} $ \cite{Liu:2006nn}, while for other type of spacetimes takes different values as expected, especially when other dimensionful scales of the theory are present \cite{Liu:2006he,Sadofyev:2015hxa,Chernicoff:2012bu,Chernicoff:2006hi,Caceres:2006ta}.

In this work we investigate further the properties of the non-local observables dependence on the Lorentz factor and how much the degree of symmetry affect the studies. Our motivation is primarily of theoretical interest and our study includes a new type of approach based on the properties of the background's asymptotic expansion, while working on the rest frame of the bound state. The holographic theories we consider have a non-trivial renormalization group flow. Their ultraviolet conformal fixed point is conformal and a Fefferman-Graham asymptotic expansion is applicable in the near boundary regime. Moreover, along the flow we allow rotational symmetry breaking,  as a result the  infrared fixed point is allowed in principle to have reduced symmetries. We point out that our approach is generic, we do not specify the holographic theory and our formalism and results are applicable for  any asymptotically AdS theory with the mentioned symmetries. Example of such specific  holographic theories with non-trivial renormalization group flows include \cite{Mateos:2011ix,Giataganas:2017koz,Hoyos:2020zeg}.

Our work can also be though to be motivated from recent developments on known strongly coupled systems and in particular of the  quark-gluon plasma.  It has been  known for a long time that the suppression in the number of $J/\Psi$ mesons \cite{NA50:2004sgj} may be related to the high transverse momenta. Moreover, recent experimental results show that certain hadrons, for example the  $\L$-hyperons, which are produced in noncentral collisions exhibit nonzero spin polarization. This  implies that the strongly coupled plasma is rotating and precise measurements suggest that it is the most vortical fluid observed to date \cite{STAR:2017ckg}. Holographic dual models that describe such fast rotating plasmas, will exhibit breaking of the rotational symmetry due to the rotation at least on a plane. Candidates of such theories are ones that have been deformed by relevant and marginal operators, which can be still asymptotically AdS, while their infrared fixed point has broken symmetry due to a rotation and they can be included in our generic formalism as special cases. The Wilson loop scaling analysis in those theories could have extra minor computational complication but there is no obvious reason that the velocity dependence on the screening length changes for a rotating motion compared to that of a linear motion. In fact in the special case of AdS, the holographic analysis for rotating plasmas supports qualitative the quarkonium suppression observed in heavy ion collisions for linear and rotating motion in the same way. For the spinning bound states in thermal backgrounds their qualitative \cite{Giataganas:2022qgq} and quantitative \cite{Peeters:2006iu} dependence of the screening length on the velocity is consistent with the one reported for linear boosts. So we expect that our classification on the scaling powers we have done in this work applies in the case of rotation as well, with potentially modified classification conditions. Recently, there is an extensive effort in holography to understand bound states and other aspects of the fast rotating strongly coupled plasmas, for example \cite{Giataganas:2022qgq,Garbiso:2020puw,Chen:2022obe,Chen:2020ath, Golubtsova:2021agl,Golubtsova:2022ldm}. It is worthy to note that the behavior of the critical temperature in presence of rotation has been investigated with different type of approaches, see for example \cite{Jiang:2016wvv,Fujimoto:2021xix,Braguta:2021jgn,Chernodub:2020qah}, where the vast majority of these studies suggests lower critical temperature for the phase transitions and easier melting of the bound states, in agreement with the holographic findings. Notice that these observations are also in agreement with computations in strongly coupled theories where the rotational symmetry is broken by other mechanisms, for example via the backreaction of a strong external field, \cite{Giataganas:2012zy,Giataganas:2017koz,Gursoy:2021efc,Giataganas:2013lga,
Giataganasimv}. This alternatively suggests the breaking of the symmetry as one of the dominant causes for the suppression on the heavy quarks and the lowering of the critical temperature of the phase transitions.

In our work we analyze analytically the quantities associated with the heavy bound state in the plasma wind for asymptotically AdS theories, and we find their scaling exponents with respect to the Lorentz factor for ultrarelativistic motion. Despite the fact that our theories are asymptotically AdS, we show that different velocity scaling powers are allowed for the Wilson loop observables. The different scalings are associated with different asymptotic background conditions we specify. As a result there is no universality in the scalings although the set of the scaling values  is finite and discrete. Once the background satisfies an asymptotic condition, a unique scaling power is determined.  The reason that a variety in scalings is possible, is that different background conditions correspond to higher order geometric contributions from the asymptotic expansion on the observables. In section 2, we rotate and boost the plasma that our dipole bound state is placed in order to set up our system in full generality. In section 3, we examine Wilson lines in general holographic boosted backgrounds,  then we apply the formalism in the asymptomatically expanded background in the ultrarelativistic limit. The observables we study for ultrarelativistic motion receive their leading contribution from the near boundary regime.    In section 4, we compute the expectation value of the Wilson loop in generic holographic boosted spacetimes. Then we take the ultrarelativistic limit and assume asymptotically AdS geometries to find how the various quantities scale with the Lorentz factor. For instance, for the screening length  we find $L_{max}\sim \prt{1-v^2}^{\ff{1}{2k}}$, where $k=1,2$ depending on the asymptotic behavior of the metric. We find the conditions of the asymptotic expansion that are associated with each scaling power $k$ and discuss their implications.

\section{Gravitational Background: Rotation and Boost}

Let us assume a space  that along the RG flow is allowed to have a potential breaking of rotational symmetry in $d$ spatial dimensions, where the $SO(d)$ boundary symmetry is broken along the flow to infrared to $SO(d-d_1)$  in the $(\vec{x_i})$ plane and $SO(d_1)$  in the $(\vec{x_j})$ plane.  To set up our system we consider  a dipole state of random orientation in the space. We examine backgrounds of only radial dependence and therefore our analysis can be reduced without any loss of generality to a three-spatial dimensional configuration with an isotropic plane $x_1 x_2$ preserved along the whole renormalization group flow  and a special direction $x_3$ ($d_1=1$) that breakes the rotational invariance in the bulk. The dipole experiences a wind of velocity $v$ that lies say  in a $x_1 x_3$-plane, and forms an angle $\theta$ with the $x_3$-direction. We align the new coordinate system at the orientation of the wind with an application of a rotation coordinate transformation of angle $\theta$  on $x_1 x_3$ plane. Then we can perform a boost along the new aligned direction of the wind that coincides with one axis to eventually obtain the rotated boosted metric
\bea\nn
ds^2= &&\tilde{g}_{00}(u) dt^2+\tilde{g}_{11}(u) d\tilde{x}_1^2+\tilde{g}_{22}(u) d\tilde{x}_2^2+\tilde{g}_{33}(u) d\tilde{x}_3^2+\tilde{g}_{uu}(u) du^2\\&&+2 \tilde{g}_{01}(u) d\tilde{x}_0 d\tilde{x}_1
+ 2  \tilde{g}_{03}(u) d\tilde{x}_0 d\tilde{x}_3+ 2\tilde{g}_{13}(u) d\tilde{x}_1 d\tilde{x}_3 ~,\la{metric1}
\eea
where
\bea\nn
&&\tilde{g}_{00}(u)= g_{00}(u) \cosh^2 \eta +\prt{g_{33}(u) c_{\th}^2 +g_{11}(u)s_{\th}^2} \sinh^2\eta~,\quad \tilde{g}_{11}(u)=  g_{11}(u) c_{\th}^2 +g_{33}(u)s_{\th}^2~, \\ \nn
&&\tilde{g}_{33}(u)= g_{00}(u)\sinh^2 \eta+\prt{g_{33}(u) c_{\th}^2 +g_{11}(u)s_{\th}^2} \cosh^2\eta~,\\\nn
&&\tilde{g}_{01}(u)=  \frac{1}{2} \prt{g_{33}(u)   -g_{11}(u) }\sinh\eta s_{2\theta}~, \quad  \tilde{g}_{13}(u)=  \frac{1}{2} \prt{g_{11}(u)   -g_{33}(u) }\cosh\eta s_{2\theta}~,  \\  \nn
&&\tilde{g}_{03}(u)= - \prt{g_{00}(u)   +g_{11}(u)\sin^2 \theta+g_{33}(u)c_{\th}^2 }\sinh\eta\cosh\eta ~.  
\eea
The tildes refer to the boosted rotated background metric and the metric elements $\tilde{g}$ that do not appear in the above expressions coincide with the initial holographic background $\tilde{g}_{ij}=g_{ij}$. The rapidity $\eta$ is given by the $\cosh^2 \eta=\g^2=1/\prt{1-v^2}$, where $\g$ is the Lorentz factor and $v$ is the velocity of the boost. The setup is particularly convenient to describe the direction of the hot wind and the dipole orientation in a generic space-time. For example, the angle $\theta=0$ corresponds to having a wind blowing along the $x_3$ direction, boosting therefore the system along the special anisotropic direction, while the angle $\th=\pi/2$ corresponds to having the wind within the transverse $x_1 x_2$ plane and the boost of the systems happens  there. When we refer to the dipole orientation is with respect to initial system before any boost and rotation. Notice that analysis is in the probe's rest frame.

\section{Velocity Laws for Wilson Lines}

Let us start by studying Wilson lines which correspond to free heavy colour charged singlets in the wind. To parametrize the string we fix it to the radial gauge while we embed the string of single boundary endpoint  $\prt{x_1(\s), x_3(\s)}$, taking into account the symmetry of the theory. The action reads
\be\la{act1}
S=-\frac{1}{2\pi\a{}'} \int d\s \sqrt{ A_0+ A_1 \tilde{x}_1{}'(\s)^2 +A_3 \tilde{x}_3{}'(\s)^2+A_{13} \tilde{x}_1{}'(\s) \tilde{x}_3{}'(\s)}~.
\ee
The functions $A_i \prt{\th,\eta,\s}$ are given by
\bea\la{als}
A_0=-\tilde{g}_{tt} \tilde{g}_{uu}~,\quad
A_1=\tilde{g}_{01}^2 -\tilde{g}_{00}\tilde{g}_{11}~,\quad A_3=\tilde{g}_{03}^2-\tilde{g}_{00}\tilde{g}_{33}~,\quad A_{13}=2\prt{\tilde{g}_{01}\tilde{g}_{03}-\tilde{g}_{00}\tilde{g}_{13}}~,
\eea
where the metric elements are of the rotated boosted of the metric \eq{metric1}.  It is straightforward to obtain the conjugate momenta $\P_1=\pp \cL/\pp \tilde{x}_{1}'$ and $\P_3 =\pp \cL/\pp{\tilde{x}_{3}'}$,  then we can solve the system of the two equations for $\tilde{x}_{1}'$ and $\tilde{x}_{3}'$ to obtain
\bea\label{p1sol}
&& \tilde{x}_{1}'^2=\ff{4 A_0 \prt{\P_3 A_{13}-2\P_1 A_3}^2}{\prt{A_{13}^2-4 A_1 A_3}\prt{4 \P_1^2 A_3-4 \P_1 \P_3 A_{13}+A_{13}^2+4 A_1\prt{\P_3^2-A_3}}}~,\\\la{p3sol}
&& \tilde{x}_{3}'^2=\ff{4 A_0 \prt{\P_1 A_{13}-2\P_3A_1}^2}{\prt{A_{13}^2-4 A_1 A_3}\prt{4 \P_1^2 A_3-4 \P_1 \P_3 A_{13}+A_{13}^2+4 A_1\prt{\P_3^2-A_3}}}~.
\eea
The requirement of real solutions for the worldsheet, suggests that the momenta remain real along the full string. This imposes the existence of a horizon on the worldsheet, at a critical value $u_c$ that is given by solving the algebraic equations requiring the numerators and the denominators of \eq{p1sol}, \eq{p3sol}  to have coinciding zeros
\be\la{wshorizon}
4 A_1 A_3=A_{13} ^2 \big|_{u=u_c}~, \qquad \P_1=\ff{A_{13} \P_3}{2 A_3} \bigg|_{u=u_c}~,\qquad  A_{13}= 2 \P_1 \P_3 \big|_{u=u_c}~.
\ee
Notice the simplicity of the above equations which also leads to a further simplification of the expressions of the momenta evaluated at $u_c$
\be\la{momsim}
\P_3^2=A_{3}\big|_{u=u_c}~,\qquad \P_1^2=\ff{A_{13}^2}{4 A_3}\bigg|_{u=u_c}~. 
\ee
It is useful to express the above conditions in terms of the initial holographic background elements, since they take a compact form
\be
 \P_3^2=-g_{00}\prt{g_{33}c_{\th}^2 +g_{11}s_{\th}^2}\big|_{u=u_c}~,\quad
 \P_1^2=\ff{-g_{00}\prt{g_{11}-g_{33}}^2 s_{2\th}^2  \cosh^2\eta}{4 \prt{g_{33}c_{\th}^2 +g_{11} s_{\th}^2}}\bigg|_{u=u_c}~.
\ee
Finally, the on-shell action comes by the substitution of \eq{p1sol} and \eq{p3sol} into the action \eq{act1} and can be written as
 \be\la{action3}
S=-\frac{1}{2\pi\a'}\int d \s\sqrt{\frac{ A_0 \prt{A_{13}^2-4A_1 A_3}}{A_{13}^2+4 A_1\prt{\P_3^2-A_3}+4 \P_1^2 A_3-4 \P_1 \P_3 A_{13}}} ~,
\ee
which is related to the energy of the system. The system of our equations \eq{p1sol} and \eq{p3sol} is solvable analytically only in certain limits, otherwise a numerical treatment is required.
Of particular interest  is the  ultra-relativistic limit, since the regime that plays a leading role is the near the boundary where we are allowed to investigate the system perturbatively.

\subsection{Worldsheet Horizon Equation and Conjugate Momenta}

The existence of the string worldsheet horizon is determined by \eq{wshorizon}, the equation can be written in the following form
\be\la{ws2}
g_{11} g_{33}\prt{g_{00}+v^2 \prt{g_{33}c_{\th}^2+g_{11}s_{\th}^2}}\cdot\prt{\prt{g_{11} s_{\th}^2 +g_{33} c_\th^2}\prt{1-v^2}}^{-1}=0\big|_{u=u_c}~.
\ee
where we include the denominator in order to comment on the asymptotic behavior of the solutions. 
This is an algebraic equation that determines the $u_c$ dependence on the velocity in the background under study.
Even without knowing the exact background the equation the behavior of the criticality in the string worldsheet can be extracted. For zero or low velocity $v\simeq0$, we have static quarks and the worldsheet critical value $u_c\simeq u_h$ approaches the black hole horizon resulting to straight string solutions. For these velocities we essentially solve \eq{ws2} for the blackening factor in $g_{00}$. The opposite ultra-relativistic limit $v\rightarrow 1$, leads the solution of eq. \eq{ws2} to the boundary regime in order to compensate the infinity generated by the denominator.

The equation \eq{ws2} can be solved for $g_{00}$
\be \la{ucsol2}
g_{00}=-v^2\prt{g_{33} c_{\th}^2 +g_{11} s_{\th}^2}\big|_{u=u_c}~,
\ee
which directly simplifies the momenta \eq{momsim}  
\be\la{mom3}
\P_1^2=\ff{\prt{g_{11}-g_{33}}^2s_{2 \th}^2 ~v^2}{4 \prt{1-v^2}}\bigg|_{u=u_c}~,\qquad \P_3^2= \prt{g_{33} c_{\th}^2 + g_{11} s_{\th}^2}^2 v^2\big|_{u=u_c}~,
\ee
where the metric elements in the right hand side have still implicit dependence on the velocity via the $u_c$ solution of the algebraic equation \eq{ucsol2}.

Nevertheless, from the equations \eq{mom3} suffices to observe that for an isotropic holographic  renormalization flow, $\P_1=0$. To understand this notice that there is a rotation to align the direction of the wind along the $\tilde{x}_3$ direction. For isotropic theories, all three directions are equivalent, which means that the non-diagonal $\tilde{g}_{xi}$ metric elements vanish and therefore only the momentum along the plasma wind is non-zero.  At the same time $\P_3$ becomes independent of $\th$ and it is never zero for the same reason, after the rotation the wind blows always along the direction $\tilde{x}_3$.   Moreover, even for a non-trivial anisotropic background when the motion happens exclusively along one of the axes $x_1$  or $x_3$ of the initial coordinate system, there is a vanishing $\P_1$  momentum along the orthogonal axis, since the direction of the wind is along  $\tilde{x}_3$. 

\subsection{Ultra-Relativistic Analysis}

As we have elaborated in the previous section  $v\rightarrow 1$ implies that $u_c \rightarrow (boundary)$. Let us assume a space which is asymptotically AdS, and along the RG flow is allowed to have a potential breaking of rotational symmetry.

\subsubsection{Asymptotically AdS Renormalization Group  Flows}

The near boundary expansion of the original theory takes the form
\be\la{expans}
ds^2=\ff{du}{u^2}+\g_{ij} dx^i dx^j~,
\ee
with
\be
\g_{ij}=\frac{1}{u^2}\prt{\eta_{ij}+u^2 g^{(2)}_{ij}+u^4 \prt{ g_{ij}^{(4)}+ h_{ij}^{(4)} \log u}+\ldots}~.
\ee
In order to investigate the expansion in detail and to investigate all the cases let us distinguish the two type of motions with respect to the $x_3$ direction.
\newline
\textbf{Motion for $\th\neq \pi/2$ }:\newline
For a motion outside the transverse plane, $\th\neq \pi/2$,  the momentum along $\tilde{x}_1$ reads directly from \eq{mom3} without the need of solving the algebraic equation for $u_c$ since it is independent of it
\be\la{finalmoma1}
\P_1^2\simeq\frac{\prt{g_{11}^{(2)}-g_{33}^{(2)}}^2\sin^2 2\th}{4(1-v^2)}~.
\ee
In contrast, the momentum along $x_3$ depends on $u_c$, so that \eq{ucsol2} needs to be solved explicitly  to give
\be\la{soluca1}
u_c^2\simeq \ff{2\prt{1-v^2}}{A_5^{(2)}}~,
\ee
where $A_5^{(2)}$ is defined via
\be\la{a5def}
A_5^{(2)}:= 2 g_{00}^{(2)}+g_{11}^{(2)}+g_{33}^{(2)}+\prt{g_{33}^{(2)}-g_{11}^{(2)}} c_{2\th}~.
\ee
Notice the $u_c\sim \sqrt{1-v^2}$ dependence which is crucial in what follows and that  the \eq{soluca1} is assumed for now to be regular. For example, if we had a fully isotropic geometry \eq{soluca1} would have been  singular. The substitution of $u_c$ in equation \eq{mom3} leads to the following  expression for the remaining momentum
\be\la{finalmoma3}
 \P_3^2=\frac{ A_5^{(2)}{}^2}{4\prt{1-v^2}^2}~.
\ee
The validity of the derived expressions rely on the condition that the expressions are regular otherwise higher order terms need to be considered, a case we examine later.   To obtain eventually the scaling of the action with the velocity, we substitute  the momenta \eq{finalmoma1}  and \eq{finalmoma3} to \eq{action3} to obtain
\be  \la{ssb1}
2\pi \a' S_{\th\neq \ff{\pi}{2}}\simeq -2  \int_0^{u_c} d u \ff{1}{u^2}\prt{1-\ff{u^2 A_5^{(2)}}{4\prt{1-v^2}}} ~.
\ee
In order to extract the Lorentz factor dependence, we  rescale $u= u_s \prt{1-v^2}^{1/2}$ so that we extract the $v$ dependence of the critical point $u_c$. Then the ultrarelativistic expansion scales with the Lorentz factor as
\be  \la{ss1}
2\pi \a' S_{\th\neq \ff{\pi}{2}}\simeq -2  \prt{1-v^2}^{-\ff{1}{2}}\int_0^{u_{sc}} d u_s \ff{1}{u_s^2}\prt{1-\ff{1}{4}u_s^2 A_5^{(2)}}%
\ee%
The action is integrated twice to count the energies of both free moving particles and the upper limit $u_{cs}$ is the rescaled worldsheet horizon. It serves as an upper bound in the integration since after this point the string is causally disconnected from the boundary. All the $v$ dependence has been extracted outside the action and we conclude that $S_{\th\neq \ff{\pi}{2}} \sim  \prt{1-v^2}^{-\ff{1}{2}}$, when $u_c$ given by the equation \eq{soluca1} is regular.

Let us examine backgrounds where the expression  \eq{soluca1} is singular to show that the leading contributions comes from higher orders in the asymptotic expansion. For this to happen the holographic background has to satisfy
 \be \la{cona1}
 A_5^{(2)}=0~,
\ee
which can occur only for a fixed angle $\th$ or for an isotropic background. Let us again work with the generic case and for simplicity we  assume that the leading leading logarithmic contributions are absent, $h_{ij}^{\prt{4}}=0$, or cancel each other in the expressions. Then the solution of the critical point of the worldsheet becomes
\be
u_c^2=\sqrt{2} \ff{\sqrt{1-v^2}}{\sqrt{A_5^{(4)}}}~,
\ee
where we have used the condition \eq{cona1} and we require the above solution to be real.  $A_5^{(4)}$ is defined in the same way as  $A_5^{(2)}$ in \eq{a5def} but with the next order terms. The leading contributions in the momenta now are found as
\be
\P_1^2=\ff{\prt{g_{33}^{(2)}-g_{11}^{(2)}}^2 s_{2\th}^2}{4\prt{1-v^2}}~, \qquad
 \P_3^2=\ff{ A_5^{(4)}}{2\prt{1-v^2}}~,
\ee
while the action takes the form
\be  \la{ss2}
S_{\th\neq \ff{\pi}{2}}\simeq   -\ff{\prt{1-v^2}^{-\ff{1}{4}}}{ 2 \pi\a'}\int_0^{u_s} d u_s\ff{1}{{u_s}^2}\prtt{1-\ff{1}{16}\prt{4 A_5^{(4)}-\prt{g_{00}^{(4)}+g_{33}^{(4)}}+\prt{1-c_{4\th}}\prt{g_{33}^{(2)}-g_{11}^{(2)}}^2}}   ~,
\ee%
where we have rescaled the radial coordinate as $u_s=u \prt{1-v^2}^{1/4}$. We point out, that only the leading terms four our study are enough to determine the scaling.

To summarize the findings so far, for motion with $\th\neq \pi/2$, the free energy scales with the Lorentz factor as $S\sim \prt{1-v^2}^{-\ff{1}{2}}$ as long as the asymptotic condition $A_5^{(2)} \neq 0$ holds, where $A_5^{(2)}$  is given by \eq{a5def}. Otherwise, if the background satisfies the above condition for $A_5^{(2)}$  asymptotically,  we obtain  $S\sim \prt{1-v^2}^{-\ff{1}{4}}$.\newline
\textbf{Motion for $\th=\pi/2$ }:\newline
For the motion in the transverse plane, $\th=\pi/2$, we deal with more compact expressions. Working again in the leading order we obtain from the solution of \eq{ucsol2} for the $u_c$:
\be\la{soluca2}
u_c^2=\ff{1-v^2}{g_{00}^{(2)} +g_{11}^{(2)}}~,
\ee
which results for the momenta
\be\la{finalmoma2}
\P_1=0~,\qquad \P_3^2=\frac{ \prt{g_{00}^{(2)} +g_{11}^{(2)}}^2 }{\prt{1-v^2}^2}~.
\ee
The action \eq{action3} in the near boundary regime comes by the use of the momenta \eq{finalmoma2} and the rescaling $u= u_s \sqrt{1-v^2}$ to isolate the velocity dependence and we obtain
\be  \la{ss3}
 \pi \a'S_{\th= \ff{\pi}{2}}\simeq -\prt{1-v^2}^{-\ff{1}{2}}\int_0^{u_{cs}} d u_s \ff{1}{u_s^2}\prt{1-u_s^2 \ff{g_{00}^{(2)}+g_{11}^{(2)}}{2}} ~.
\ee

The leading contribution changes when the equation \eq{soluca2} is singular for the holographic background. Then the leading order to the Wilson lines will be the next one in the asymptotic expansion.  In this case the background satisfies 
\be\la{cona2}
g_{00}^{(2)} +g_{11}^{(2)}=0~,
\ee
and the worldsheet horizon is found by solving  \eq{ucsol2} and is equal to
\be\la{u2}
u_c^2\simeq  \ff{\sqrt{1-v^2}}{\sqrt{g_{00}^{(4)} +g_{11}^{(4)}}}~.
\ee
The above expression is valid even with logarithmic terms present, as long as $h_{00}^{(4)}=h_{11}^{(4)}$. However, to simplify the presentation let us consider vanishing logarithmic contributions.
The momentum $\Pi_3$ takes the simple form
\be \la{finalmom3}
\P_3^2\simeq \ff{g_{00}^{(4)} +g_{11}^{(4)}}{1-v^2}~.
\ee
The Lorentz scaling in the action  can be found by rescaling $u= u_s \prt{1-v^2}^{1/4}$ where we obtain in the near boundary regime the divergent action as
\be \la{ss4}
 \pi \a' S_{\th= \ff{\pi}{2}}\simeq -\prt{1-v^2}^{-\ff{1}{4}}\int_0^{u_{cs}} d u_s \ff{1-u_s^4\prt{g_{00}^{(4)}+g_{11}^{(4)}}}{u_s^2\prt{1-2 u_s^2 \prt{g_{00}^{(4)}+g_{11}^{(4)}}} }~.
\ee
In summary we find from equations \eq{ss1}, \eq{ss2}, \eq{ss3} and \eq{ss4}, that $S$ scales as $S\sim \prt{1-v^2}^{-\ff{1}{2k}}$, where $k=1,2$ and depends on the angle of motion and the asymptotic behavior. When the leading terms of the asymptotic expansion of the background satisfy \eq{cona1} and \eq{cona2} then $k=2$ and the scalings read off the equation \eq{ss2} and \eq{ss4}, otherwise $k=1$ and the actions is given asymptotically by  \eq{ss1} and \eq{ss3}.

\section{Velocity laws for Wilson loops and  the Screening Length}

In the previous section we have computed the leading contributions of the velocity for the singlets' motion. In this section we compute the expectation value of the Wilson loop from where we read off the velocity dependence for the heavy bound state potential and for the screening length. We will apply the mass renormalization scheme, subtracting the free energy divergences of the bound state from the infinite masses of the singlets we derived in the previous section in order to extract the interaction energy of the bound state. We take the orientation of the dipole to be arbitrary with respect to the velocity of the plasma and the $\tilde{x}_i$ coordinates. For the Wilson loop computation we fix the radial gauge and have the string embedded along the spatial directions $\prt{x_1(u),x_2(u),x_3(u)}$. The string has a turning point at $u_0$, and the action reads
\be\la{act2}
S=-\frac{1}{2\pi\a{}'} \int d\s \sqrt{ A_0+ A_1 \tilde{x}_1{}'(\s)^2+ A_2 \tilde{x}_2{}'(\s)^2 +A_3 \tilde{x}_3{}'(\s)^2+A_{13} \tilde{x}_1{}'(\s) \tilde{x}_3{}'(\s)}~,
\ee
where the functions $A_i \prt{\th,\eta,\s}$  are given by \eq{als} and additionally
\bea
A_2=-\tilde{g}_{00}\tilde{g}_{22}~.
\eea
We can solve the system of the momenta  $\P_1=\pp \cL/\pp \tilde{x}_{1}'$,  $\P_2=\pp \cL/\pp \tilde{x}_{2}'$ and $\P_3 =\pp \cL/\pp{\tilde{x}_{3}'}$ with respect to the derivatives of the coordinates to get
\bea\nn
&&\tilde{x}_1' =\ff{2 \sqrt{A_0 A_2}\prt{2\P_1 A_3-\P_3 A_{13} }}{\sqrt{-\prt{A_{13} ^2- 4 A_1 A_3}\cD}}~,\\ \la{x1}
&&\tilde{x}_2'=\ff{\P_2 \sqrt{-A_0\prt{A_{13}^2-4 A_1 A_3}}}{\sqrt{A_2 \cD}}~,\\ \nn
&&\tilde{x}_3'=\ff{ 2\sqrt{A_0 A_2}\prt{2\P_3 A_{1}-\P_1 A_{13}}}{\sqrt{-\prt{A_{13} ^2- 4 A_1 A_3}\cD}}~,
\eea
with a common denominator $\cD$ which reads
\be\la{dd1}
\cD=A_{13}^2\prt{\P_2^2-A_2} +4 \P_1 \P_3 A_{13} A_2-4\prt{\P_1^2 A_2 A_3 +A_1\prt{A_2\prt{\P_3^2-A_3}+\P_2^2 A_3}}~.
\ee
The boundary length of the state is
\be \prt{ L_1, L_2, L_3}=\prt{L s_{\th_d} s_{\phi_d}, L s_{\th_d} c_{\phi_d}, L c_{\th_d}}~,
\ee
where $\prt{\th_d, \phi_d}$ are the bound state orientation angles. To determine it we need to integrate the equations \eq{x1} as
\be\la{lls}
L_i =2\int_0^{u_0}  \tilde{x}_i ~d\s~. 
\ee
By substituting the equations of motion \eq{x1} to the action \eq{act2} we can eliminate the derivatives and eventually  express it in terms of the momenta
 \be\la{actionbs3}
S=-\frac{1}{\pi\a'}\int_0^{u_0} d \s\sqrt{\frac{ - A_0  A_2 \prt{A_{13}^2-4A_1 A_3}}{\cD}} ~.
\ee
Our aim is to extract the dependence behavior of the interaction energy \eq{actionbs3} on the Lorentz factor for non-trivial holographic renormalization group flows. The turning point of the string worldsheet is specified by solving the algebraic equation $\cD=0$ given by the \eq{dd1}. The algebraic equation  simplifies considerably for high and low velocities. At the limit of high velocities where our primary interest is, it reads
\be
\ff{g_{22}\prt{g_{00}+g_{33} c_\th^2+g_{11} s_\th^2 }\prtt{g_{11}\prt{\P_3^2 g_{33}+g_{00}^2
		g_{33}+g_{00}\prt{\P_3^2+g_{33}^2}c_\th^2}+g_{00} g_{33}\prt{\P_3^2 +g_{11}^2}s_\th^2}}{\prt{1-v}^2}=0,
\ee
with a solution for the turning point $u_0$ that approaches the near boundary regime. So far, we have not made any approximations or assumptions and all the expression derived hold for  any holographic background. To progress further so as to find the explicit velocity dependence, we apply the formalism on non-trivial RG holographic flows that are asymptotically AdS. 
Let us split the analysis again with respect to the direction of the wind  and the values of angles $\th$.

\subsection{Asymptotically AdS Renormalization Group Flows}
\textbf{Motion for $\prt{\th\neq \pi/2}$}\newline
The state's boundary length integrals \eq{lls} in the asymptotic regime of the metric \eq{expans}, after some algebra take the compact form
\bea \la{ll2}
L_i\simeq 2\int_0^{u_0} du ~\ff{\P_i u^2}{\sqrt{D_0}}\prt{1+ u^2 \prt{s_i \ff{A_5^{(2)}}{4\prt{1-v^2} } +\tilde{s}_i\ff{\prt{g_{33}^{(2)}-g_{11}^{(2)}}s_{2\th} }{2\sqrt{1-v^2}}}}~,
\eea
where $s_1=s_2=s_3/3=-1$ and $\tilde{s}_1=\tilde{s}_3^{-1}=\P_3/\P_1~, ~\tilde{s}_2=0~$ and the $u-$independent $A_5^{(2)}$ has been defined in \eq{a5def}. Notice that in this section we omit terms in the expansion that will be proven to be subleading with the rescalings below.  $\cD_0$ is defined through the  asymptotic expansion of $\cD$ as
\be\la{d1exp}
\cD\simeq \ff{4}{u^{12}}\prtt{1-\ff{ A_5^{(2)}}{1-v^2} u^2+\prt{\ff{ A_5^{(2)}{}^2}{4\prt{1-v^2}^2}- \sum_{i=1}^{3}\P_i^2+ \ff{C_1}{1-v^2}+C_2}u^4 }:=\ff{4}{u^{12} }\cD_0~,
\ee
where $C_i$ are u-independent terms that will be proven subleading in the subsequent analysis.
The action for the bound state can be found from \eq{actionbs3} by putting all the expansions together and reads
\be \la{actionb3}
S\simeq-\frac{1}{\pi\a'} \int_0^{u_0} d u\ff{1}{u^2 \sqrt{D_0}}\prt{1 - \ff{3A_5^{(2)} }{4\prt{1-v^2}} u^2}~.
\ee
At this stage we have to impose the natural requirement, \cite{Maldacena:1998im,Drukker:1999zq,Chu:2008xg,Giataganas:2022qgq}, to have an action with leading divergences that are canceled by the ultraviolet divergence of the infinite massive singlets \eq{ss1}, and that the momenta increase with the increase of the velocity.  To extract the correct scaling dependence, we rescale $u$ and $\Pi$ at the ultrarelativistic limit as $u= u_s \sqrt{1-v^2}$ and $\P_i=\P_{s,i} \prt{1-v^2}^{-1}$.  The energy now becomes in the rescaled $u_s$ coordinates
\be \la{actionbsa3}
S\simeq -\frac{1}{\pi\a'} \ff{1}{\sqrt{1-v^2}}\int_0^{u_{s0}} d u_s \ff{1}{u_s^2\sqrt{D_{s0}}}\prt{1-  \ff{3A_5^{(2)} }{4}u_s^2}\simeq -\frac{1}{\pi\a'} \ff{1}{\sqrt{1-v^2}}\int_0^{u_{s0}} d u_s \prt{\ff{1}{u_s^2} + \ff{A_5^{(2)} }{4}}~
\ee
and boundary lengths read from equation \eq{ll2}
\be \la{llf}
L_i\simeq 2 \prt{1-v^2}^\ff{1}{2} \P_{s,i}\int_0^{u_{s0}} d u_s~ \ff{u_s^2}{\sqrt{D_{s0}}}\prt{1+ s_i \ff{u_s^2 A_5^{(2)}}{4}}~,
\ee
where notice that $D_{s0}$ is obtained from the rescaling in $D_0$ of \eq{d1exp} and is $v$-independent
\be
D_{s0}=1- A_5^{(2)} u_s^2+\prt{\ff{ A_5^{(2)}{}^2}{4}- \sum_{i=1}^{3}\P_{s,i}^2 }u_s^4~.
\ee
The action \eq{actionbsa3} indeed has the same ultraviolet asymptotics  with the Wilson line action \eq{ss1} and the divergences cancel each other as expected.  

Notice that the integrands of $L_i$ in equations \eq{llf} are independent of the Lorentz factor and therefore $L_i\sim (1-v^2)^{1/2}$. The strings under study have a maximum length that depends on the scales of the theory and in the ultrarelativistic limit this is obtained by finding the maximum possible distance that the bound state exist, merely  by the maximization of \eq{llf}. The velocity dependence has extracted outside the integrands in   \eq{llf} and the integrals are bounded. Therefore  the maximum separation between a bound quark-antiquark state, which is the screening length for the bound state, scales as $L_{max}\sim \prt{1-v^2}^{\ff{1}{2}}~.$

Let us assume that there exist a scenario for a certain angle that the condition $A_5^{(2)}=0$, like the prior analysis of the singlet \eq{cona1}.  We expect that the leading action divergences  scales with the Lorentz factor as in \eq{ss2}.  The asymptotic expansion of $\cD$ becomes
\be
\cD\simeq \ff{4}{u^{12}}\prtt{1+u^4\prt{-\ff{  A_5^{(4)}}{\prt{1-v^2}}- \sum_{i=1}^{3}\P_i^2  +\ff{C_1}{1-v^2}+C_2}}=\ff{4}{u^{12} }\cD_0~,
\ee
where the $C_i$ terms will be proved subleading and are not the same as in previous expressions where we have used the same notation. The boundary length now reads
\bea \la{ll3}
L_i\simeq 2\int_0^{u_0} du ~\ff{\P_i u^2}{\sqrt{D_0}}\prt{1 +\tilde{s}_i u^2 \ff{\prt{g_{33}^{(2)}-g_{11}^{(2)}}s_{2\th} }{2\sqrt{1-v^2}}+u^4 \prt{s_i\ff{ A_5^{(4)}}{4\prt{1-v^2}}+\hat{s}_i \ff{\prt{g_{33}^{(2)}-g_{11}^{(2)}}^2\prt{c_{4\th} -1}}{8\prt{1-v^2}}}}~,
\eea
where $\hat{s}_1=\hat{s}_2=0~,~ \hat{s}_3=1$ and the subleading terms will be proven that do not contribute. The action takes the form
\be \la{actionb4}
S\simeq-\frac{1}{\pi\a'} \int_0^{u_0} d u\ff{1}{u^2 \sqrt{D_0}}\prt{1 - \ff{3 A_5^{(4)}}{4\prt{1-v^2}} u^4}~.
\ee
Acting as in previous cases, we rescale $u= u_s \prt{1-v^2}^{1/4}$ and then $\P_i$ needs to be rescaled as  $\P_i=\P_{s,i} \prt{1-v^2}^{-1/2}$ to give the energy in the new redial coordinates
\be \la{actionb33}
S\simeq -\frac{1}{\pi\a'} \ff{1}{\prt{1-v^2}^{\ff{1}{4}}}\int_0^{u_{s0}} d u_s \ff{1}{u_s^2\sqrt{D_{s0}}}\prt{1-  \ff{3  A_5^{(4)}}{4}u_s^2}~.
\ee
where
\be
D_{s0}=1- A_5^{(4)}u_s^4- \sum_{i=1}^{3}\P_{s,i}^2 u_s^4 ~.
\ee
The leading contributions to the expression of $L_i$ close to the boundary become
\be
L_i\simeq 2\int_0^{u_{s0}} du_s ~\ff{\P_{s,i} u_s^2}{\sqrt{D_{s0}}}\prt{1 +u_s^4 \prt{s_i\ff{A_6}{4}+\hat{s}_i \ff{\prt{g_{33}^{(2)}-g_{11}^{(2)}}^2\prt{c_{4\th} -1}}{8}}}~,
\ee
and therefore follows that $L_{max}\sim  \prt{1-v^2}^{\ff{1}{4}}~$ provided that the background satisfies $A_5^{(2)}=0$ for a motion along a certain angle and that the contributions of the logarithmic terms in the asymptotic expansion cancel each other or vanish.\newline
\textbf{Motion for $\prt{\th=\pi/2}$}\newline
For motion within the transverse plane, $\th=\pi/2$, the computation is along the same lines. The asymptotic expansion of $\cD$ is
\be\la{d1exp11}
\cD\simeq \ff{4}{u^{12}}\prtt{1-\ff{ 2\prt{g_{00}^{(2)}+g_{11}^{(2)}}}{1-v^2} u^2+\prt{\ff{ \prt{g_{00}^{(2)}+g_{11}^{(2)}}^2}{\prt{1-v^2}^2}- \sum_{i=1}^{3}\P_i^2+\ff{C_1}{1-v^2}+C_2} u^4 }=\ff{4}{u^{12} }\cD_0~.
\ee
The state's boundary length integrals \eq{lls} in the asymptotic regime of \eq{expans}, after some algebra take the compact form
\bea \la{llb}
L_i\simeq 2\int_0^{u_0} du ~\ff{\P_i u^2}{\sqrt{D_0}}\prt{1- s_i u^2 \ff{\prt{g_{00}^{(2)}+g_{11}^{(2)}}}{4\prt{1-v^2} } }~,
\eea
where $s_1=s_2=s_3/3=1$ as defined earlier.  The action for the bound state reads
\be \la{actionbb3}
S\simeq-\frac{1}{\pi\a'} \int_0^{u_0} d u\ff{1}{u^2 \sqrt{D_0}}\prt{1 - \ff{3\prt{g_{00}^{(2)}+g_{11}^{(2)}} }{2\prt{1-v^2}} u^2+\ff{3\prt{g_{00}^{(2)}+g_{11}^{(2)}}^2 }{2\prt{1-v^2}^2} u^4}~.
\ee
We require for the action's ultraviolet-divergences to get cancelled with the divergences  of the infinite massive singlets \eq{ss3}. In the ultrarelativistic limit we rescale $u$ and $\Pi_i$,  $u= u_s \sqrt{1-v^2}$ and $\P_i=\P_{s,i} \prt{1-v^2}^{-1}$ to get the  action for the bound state in new coordinates
\be
S\simeq-\frac{1}{\pi\a'} \ff{1}{\sqrt{1-v^2}}\int_0^{u_{s0}} d u_s\ff{1}{u_s^2 \sqrt{D_{s0}}}\prtt{1 - \ff{3}{2}\prt{\prt{g_{00}^{(2)}+g_{11}^{(2)}}  u_s^2+\prt{g_{00}^{(2)}+g_{11}^{(2)}}^2  u_s^4}}~,
\ee
extracting the right velocity dependence, where $D_{s0}$ in the rescaled coordinates reads
\be
D_{s0}= 1- 2\prt{g_{00}^{(2)}+g_{11}^{(2)}} u_s^2+\prt{g_{00}^{(2)}+g_{11}^{(2)}}^2u_s^4- \sum_{i=1}^{3}\P_{s,i}^2 u_s^4 ~.
\ee
The boundary lengths in the rescaled radial coordinate are equal to
\bea \la{llbf}
L_i\simeq 2\prt{1-v^2}^{\ff{1}{2}}\int_0^{u_{s0}} du_s ~\ff{\P_{i,s} u_s^2}{\sqrt{D_{s0}}}\prt{1- \ff{1}{4}s_i u_s^2 \prt{g_{00}^{(2)}+g_{11}^{(2)}}} ~.
\eea
Therefore the screening length of the bound state scales as $L_{max}\sim \prt{1-v^2}^{\ff{1}{2}}$
where we again observe the half-integer power.

When the leading contributions come from the next term of the FG expansion we get a different power scaling.  This happens when the asymptotic expansion of the metric satisfies $g_{00}^{(2)}=-g_{11}^{(2)}$. In this case $\cD$ takes the form
\be\la{d1exp22}
\cD\simeq \ff{4}{u^{12}}\prtt{1+\prt{-\ff{ 2\prt{g_{00}^{(4)}+g_{11}^{(4)}}}{\prt{1-v^2}}- \sum_{i=1}^{3}\P_i^2 +C_2}u^4}=\ff{4}{u^{12} }\cD_0~,
\ee
and $L_i$ reads
\bea \la{llbk}
L_i\simeq 2\int_0^{u_0} du ~\ff{\P_i u^2}{\sqrt{D_0}}\prt{1- u^2 c_i+\ff{k_i}{\prt{1-v^2}}} ~,
\eea
where $k_i$ and $c_i$ are u-independent terms. As has been mentioned earlier we omit terms that will be proven that do not affect the scaling. The action computation gives
\be \la{actionbb4}
S\simeq-\frac{1}{\pi\a'} \int_0^{u_0} d u\ff{1}{u^2 \sqrt{\cD_0}}\prt{1 - \ff{3\prt{g_{00}^{(4)}+g_{11}^{(4)}} }{2\prt{1-v^2}} u^2}~.
\ee
To extract the right asymptotic of the divergences matching the Wilson line action \eq{ss4} we rescale the radial coordinate and the momenta as $u= u_s \prt{1-v^2}^{1/4}$ and $\P_i=\P_{s,i} \prt{1-v^2}^{-1/2}$.
 The action for the bound state becomes
 \be \la{actionbb4b}
 S\simeq-\frac{1}{\pi\a'} \ff{1}{\prt{1-v^2}^{1/4}}\int_0^{u_{s0} }d u_s\ff{1}{u_s^2 \sqrt{\cD_{s0}}}\prt{1 - \ff{3\prt{g_{00}^{(4)}+g_{11}^{(4)}} }{2} u_s^2}~, \ee
 and the boundary length of the bound state scales as
\bea \la{llbbbb}
L_i\simeq 2\prt{1-v^2}^\ff{1}{4}\int_0^{u_{s0}} du_s ~\ff{\P_{s,i} u_s^2}{\sqrt{D_{s0}}} ~,
\eea
where  the leading contributions of $D_{s0}$ are $v$-independent. Therefore with the same arguing we extract the dependence of the screening length to the bound state $L_{max}\sim \prt{1-v^2}^{\ff{1}{4}}$  for $\th=\pi/2$ and  $g_{00}^{(2)}=-g_{11}^{(2)}$, assuming that the logarithmic terms of the expansion vanish.

In summary, we find
\bea \la{llfff}
L_{max}\sim \prt{1-v^2}^{\ff{1}{2k}}
\eea
where $k=1,2$ depending on the direction of motion and the asymptotic behavior of the metric. $k=2$ when the background conditions \eq{cona1} and \eq{cona2} are asymptotically satisfied for the corresponding direction of motion. Our analysis shows that theories dual to non-trivial RG flows which are asymptotically AdS and have a non-trivial renormalization group flow with broken rotational symmetry along the flow, are allowed to have bound states with a non-universal scaling dependence on the velocity, which we explicitly specify.

\textbf{Acknowledgments:}
The research work of D.G. is supported by the National Science and Technology Council (NSTC) of Taiwan with the Young Scholar Columbus Fellowship grant 110-2636-M-110-008.

\bibliographystyle{bibhep}

\providecommand{\href}[2]{#2}\begingroup\raggedright\endgroup

\end{document}